# Nanoscale brittle-to-ductile transition of the $C$15 $CaAl_2$ Laves phase

Anwesha Kanjilal[1], Ali Ahmadian[1,a)], Martina Freund[2], Pei-Ling Sun[2], Sandra Korte-Kerzel[2], Gerhard Dehm[1,*] and James P. Best[1,*]

[1] *Max-Planck-Institut für Eisenforschung GmbH, 40237 Düsseldorf, Germany*
[2] *Institut für Metallkunde und Materialphysik, RWTH Aachen University, 52056 Aachen, Germany*
[a)] *Present address: Karlsruhe Institute of Technology, Institute of Nanotechnology, 76433 Eggenstein-Leopoldshafen, Germany*
[*] *Corresponding authors: j.best@mpie.de and dehm@mpie.de*

**Abstract:**

The influence of temperature on the deformation behaviour of the $C$15 $CaAl_2$ Laves phase, a key constituent for enhancing the mechanical properties of Mg alloys up to service temperatures of 200 °C, remains largely unexplored. This study presents, for the first time, the nanoscale brittle-to-ductile transition (BDT) of this intermetallic phase through *in situ* testing including nanoindentation, scratch testing, and micropillar splitting conducted at elevated temperatures. By correlating observations from these techniques, changes in deformation of $CaAl_2$ were identified in relation to temperature. High-temperature nanoindentation quantitatively determined the temperature range for the BDT, and revealed that $CaAl_2$ undergoes a BDT at ~0.55$T_m$, exhibiting an intermediate region of microplasticity. A noticeable decrease in nanoindentation hardness was observed at ~450-500 °C, accompanied by an increase in residual indent size, while indentation cracking was not observed above 300 °C. Results from high-temperature micropillar splitting revealed cracking and brittle pillar splitting up to 300 °C, with an increase in apparent fracture toughness from 0.9 ± 0.1 MPa·√m to 2.8 ± 0.3 MPa·√m, and subsequent crack-free plastic deformation from 400 °C. Transmission electron microscopy analysis of the deformed material from nanoindentation revealed that the BDT of $CaAl_2$ may be attributed to enhanced dislocation plasticity with increasing temperature.

# 1 Introduction

Laves phases constitute one of the largest class of intermetallic compounds and are usually present as fine precipitates in various engineering alloys such as Mg alloys, steels, Ni-based superalloys, etc. [1–3]. They have been found to be important for improving the strength and high-temperature creep resistance of structural materials [1]. Laves phases are categorised into three polytypes depending on their crystal structure – cubic $C$15, hexagonal $C$14, and dihexagonal $C$36 type [1,4]. These crystals have topologically close packed structures with very high space filling, and therefore possess a high barrier to plastic deformation by dislocation motion at room temperature [1,5–7]. Hence, Laves phases are generally brittle under ambient conditions [1,5]. Additionally, and similar to other intermetallics [8,9], Laves phases can undergo a brittle-to-ductile transition (BDT) accompanied with a decrease in strength as the temperature increases [4]. A precise knowledge of the BDT temperature (BDTT) is important for Laves phase-strengthened alloys, in order to predict their high-temperature mechanical response for design purposes.

A common intermetallic precipitate present in binary Mg-Al alloys is $Mg_{17}Al_{12}$ ($T_m$ = 458 °C [10], where $T_m$ is the melting temperature), which undergoes a marked decrease in hardness from ~3.5 GPa to ~0.7 GPa at 150 °C [11]; a much lower temperature than the desired service temperature of 175–200 °C for Mg alloys [12]. The $C$15 $CaAl_2$ Laves phase is commonly found in Mg-rich ternary alloys containing Al and Ca. It has garnered interest for its potential to withstand higher service temperatures compared to other intermetallics in Mg-Al alloys, such as $Mg_{17}Al_{12}$, attributed to its high melting point of 1079 °C [13,14]. Tuning the Al-Ca ratio in ternary Mg-Al-Ca alloys can therefore favour the formation of higher melting point intermetallics such as the $C$14 $CaMg_2$ and $C$15 $CaAl_2$ Laves phases [15–17], leading to improved high-temperature mechanical properties. Nevertheless, the $CaMg_2$ phase has been noted for its inadequate creep properties attributed to its hexagonal structure [14]. This underscores the need for a detailed investigation of the high-temperature deformation behaviour of the $C$15 $CaAl_2$ Laves phase. Recent nano- and micromechanical studies have shown that $C$15 $CaAl_2$ possesses high hardness and strength [18], but a low fracture toughness at room temperature [19], typical of Laves phases. *In situ* microcantilever fracture tests on single-phase cast specimens of $C$15 $CaAl_2$ revealed a fracture toughness of ~2 $MPa\sqrt{m}$, with {110} and {112} as preferential cleavage planes [19]. Freund *et al.* performed nanoindentation and micropillar compression of the same cast $CaAl_2$ material, and reported a room temperature hardness of ~4.8 GPa and yield stress ~2–3 GPa [18]. In that study, it was possible to introduce plasticity in the brittle $CaAl_2$ during microcompression at room temperature due to the small length-scales involved, which favours dislocation plasticity over crack initiation as commonly observed for brittle materials [20–22]. While there is some understanding of the micromechanical properties of $CaAl_2$ at room temperature, knowledge regarding its elevated temperature deformation and fracture behaviour, and potential BDT is currently limited. In the sole study to the authors' knowledge investigating the high-temperature deformation of $CaAl_2$, Rokhlin *et al.* reported from microhardness tests that $CaAl_2$ was the hardest among the intermetallics in the ternary Mg-Al-Ca system, where the microhardness was observed to decrease ~20 % between 20 and 300 °C [14]. However, detailed insights into temperature-induced deformation changes, such as cracking and slip activity, or a possible BDT, were not discussed. Furthermore, the microhardness measurements in Ref. [14] were conducted on $CaAl_2$ precipitates embedded within the Mg alloy, where the compliance of the surrounding matrix could also influence the obtained results.

High-temperature deformation and the BDT, however, have been extensively investigated in high melting-point transition metal Laves phases such as $NbCo_2$, $FeCo_2$, $NbCr_2$, or $TaCo_2$ using hardness testing [23] or macroscale compression testing [4,24,25]. These studies revealed that while the intermetallics exhibited brittleness at room temperature, they experienced a decrease in hardness or significant reduction in yield strength (at 0.2% strain), accompanied by an increase in creep deformation above the BDTT of ~0.65 $T_m$. Additionally, the presence of pre-existing defects in the cast specimens, such as micro-cracks and pores, was found to limit ductility and adversely affect plasticity and fracture behaviour during macroscale testing, thereby impeding the accurate determination of the BDTT. Small-scale testing methods may offer a viable alternative to overcome this limitation for determining the BDTT of Laves phases.

Advances in nano- and micromechanical testing have opened up new possibilities to investigate small-scale mechanical properties across a wide range of temperatures, without the material being influenced by fabrication defects such as pre-existing flaws which affect fracture measurement [26]. Notched microcantilever bending and micropillar splitting techniques [27] have been used to determine the temperature-dependent fracture toughness of ceramic coatings [28,29] and BDT behaviour at small length-scales for Si [30,31] and W [32]. Nanoindentation has been widely used to study the elevated temperature mechanical properties for a variety of brittle materials and monitor plasticity changes with temperature [26]. Complementary to indentation testing, scratch testing has also been employed to probe the plasticity and cracking behaviour at small length-scales [33,34], and to determine orientation dependent changes in brittle and ductile deformation at room temperature due to changes in crack formation and scratch depth with crystal orientation [35,36], suggesting this technique to be a viable method for estimating the BDTT at small-scale. In the context of Laves phases, indentation-based testing has been employed to examine the high-temperature mechanical properties of $C$14 $CaMg_2$ [37], where the indentation hardness decreased above 0.59 $T_m$. High-temperature nanoindentation and micropillar compression were also used to determine the temperature-dependent hardness and critical resolved shear stress (CRSS) of the activated slip systems in $C$14 $CaMg_2$ between 25–250 °C (0.53 $T_m$) [38], where no change in the mechanical properties was observed within the temperature range. These studies suggest the versatility of small-scale testing methods to obtain meaningful fracture and plasticity properties of brittle materials over a wide range of temperatures, without the limitation of premature fracture induced by pre-existing flaws in the material.

In this context, the present study aims to determine the elevated temperature deformation behaviour and BDT of a cast single-phase $C$15 $CaAl_2$ Laves phase intermetallic specimen using a nano- and micromechanical testing approach. For this, high-temperature scratch testing using a commercial *in situ* nanoindenter is used as a screening method to obtain initial preliminary information on the BDT of $CaAl_2$, which is then followed by more precise quantitative determination of the BDTT from changes in hardness and fracture toughness using high-temperature nanoindentation and micropillar splitting, respectively. The mechanical properties measured at various temperatures are then correlated with their corresponding structural changes concerning crack formation and slip activity through subsequent examination of the deformed specimens using scanning and transmission electron microscopy.

## 2 Materials and methods
### 2.1 *Sample fabrication and microstructural characterisation*

Bulk single-phase specimens of the $C$15 $CaAl_2$ Laves phase were prepared by casting individual high purity elements of Al (99.999% purity, HMV Hauner GmbH & Co. KG, Germany) and Ca (98.8% purity, Alfa Aesar, Germany) in a cylindrical mould. The cast sample was then annealed at 600 °C for 24 h in an Argon atmosphere to homogenise the microstructure and relieve any residual stresses from the casting process. Small pieces of the specimen were machined using electric discharge machining and metallographically polished from 4000 grit

SiC paper down to 40 nm colloidal silica to obtain a good surface finish. Further details of the sample preparation procedure can be found in Ref. [39]. Composition analysis of the same $CaAl_2$ Laves phase material was performed using electron probe microanalysis in Ref [19]. **Fig. S1** in the Supplementary Information shows the representative microstructure of the annealed $C$15 $CaAl_2$ Laves phase where the grains had the $CaAl_2$ phase with the desired target composition of 67 at.% Al and 33 at.% Ca, with a skeleton network of $CaAl_4$ phase formed along the grain boundaries (chemical analysis of the respective phases from electron probe microanalysis is detailed in Ref. [19]). The grain orientation was characterised using electron backscattered diffraction (EBSD) performed with an EDAX system equipped with Hikari CCD camera mounted inside Zeiss Auriga dual beam focused ion beam/scanning electron microscope (FIB/SEM) and data collection was done using TSL OIM v7 software.

*2.2 High-temperature mechanical testing*

The effect of temperature on the deformation of $CaAl_2$ Laves phase was investigated using *in situ* high-temperature nano- and micromechanical testing. First, high temperature scratch testing was used to obtain preliminary insight into qualitative changes in scratch deformation morphology with temperature. Subsequently, more exact measurements of BDT and high-temperature fracture behaviour of $CaAl_2$ was obtained from *in situ* nanoindentation and micropillar splitting. All *in situ* mechanical tests were performed inside a Zeiss Gemini 500 SEM under high vacuum conditions ($< 5 \times 10^{-6}$ mbar) using a Hysitron PI-88 (Bruker, USA) testing system equipped with high-temperature module. Simultaneous heating of the indenter tip and sample was used to ensure good temperature matching and minimise thermal drift. The tip holder was affixed to a small ceramic sleeve to minimise heat loss and improve thermal insulation. For each type of test, the $CaAl_2$ specimen was positioned on the sample heating stage and fixed using three Mo pins. The sample and indenter tip temperatures were measured independently using two separate thermocouples, in close contact with the tip and sample heater for all tests.

*2.2.1 High-temperature scratch testing*

An EBSD map over the region of interest was first made, followed by indentation scratch testing with a diamond Berkovich tip (Synton-MDP AG, Switzerland) performed from room temperature to 500 °C. Scratching was performed through lateral manipulation of the sample stage over ~25 µm, with initial load of 5 mN and targeted final load of 25 mN. The natural misalignment of the sample surface relative to the lateral movement direction allowed for the study of effect of an increasing load on the deformation characteristics of the scratched material. After testing, the scratches were imaged in the SEM to correlate qualitative changes in the morphology of scratches with temperature.

*2.2.2 High-temperature nanoindentation*

High-temperature nanoindentation was performed within a single grain of the CaAl₂ intermetallic with ~(5 3 6) surface plane, determined *a priori* using EBSD. A high temperature diamond cube corner tip (Synton-MDP AG, Switzerland) was used in order to facilitate investigation into crack formation as a function of temperature. Prior to testing, the tip area function calibration was carried out on fused quartz. The nanoindentation tests were performed at eight different temperatures from 25 to 580 °C, with an average of 7 to 8 indents at each temperature. This was followed by another set of indentations made at 25 °C after cooling the sample down from the maximum temperature to check for repeatability before and after the heating excursion. Care was taken to ensure that the indents were made inside the grain, away from grain boundaries, and the spacing between the indents was maintained at greater than three times the lateral dimension of the indents to avoid the influence of overlapping plastic zones between adjacent indents [40]. The tests were performed in load-controlled mode, wherein the load was ramped up to a maximum value of $P_{max}$= 9 mN followed by holding at the peak load for 10 s (or 30 s at temperatures higher than 450 °C to minimise creep effects during unloading). A drift correction segment of 30 s was introduced during unloading at $0.1P_{max}$. The drift correction load was low enough to ensure that only linear drift occurred without any time-dependent creep deformation. Thermal drift rate was calculated from the drift correction segment for every indentation test at each temperature and the drift corrected load-displacement curves were subsequently obtained. The indentation hardness *H* was calculated for each case as [41]:

$$H = \frac{P_{max}}{A} \quad (1)$$

where the contact area *A* depends on the contact depth of penetration according to the calibrated area function of the cube corner indenter tip. Post-nanoindentation, the residual indent impressions were observed in the SEM in secondary electron mode to visualise changes in the indent size and microstructural features such as cracks and slip lines around indents at different temperatures.

*2.2.3   High-temperature micropillar splitting*

Micropillar splitting was performed to determine the effect of temperature on the fracture toughness of $CaAl_2$ Laves phase. In total, 17 cylindrical micropillars were fabricated by FIB milling in a Zeiss FIB-SEM. The final milling was performed at 30 kV, 3 nA $Ga^+$ ion beam current down to a final pillar diameter of 10 µm and aspect ratio ~1.4. The dimensions of the pillars (i.e. the pillar diameter and height) were measured after FIB milling. An overview image of the micropillar and the surface orientation of the grain are given later in **Section 3.3**. Recently, Lauener *et al.* [30] reported that for pillars with diameter greater than 10 µm, ion beam damage has limited influence on the pillar splitting fracture toughness [28,29]. The effect of ion beam and pillar size on the room temperature fracture toughness of $CaAl_2$ has also been investigated by the authors [19], where no observable differences between the apparent fracture toughness of 10 µm diameter pillars milled with either $Ga^+$ or $Xe^+$ ions were observed. Furthermore, in the same study, $CaAl_2$ micropillars of 5 and 10 µm diameter fabricated by $Ga^+$ FIB were found to have similar toughness values [19]. Hence, the pillar size and fabrication procedure used in this study will not lead to ion beam induced damage effects on toughness measurement. To facilitate fracture of the micropillars and also maintain consistency with the nanoindentation tests, the pillar splitting tests were also performed using a high temperature diamond cube corner indenter (Synton-MDP AG, Switzerland) at 25, 150, 300, 400 and 500 °C. Pillar splitting tests were conducted in pseudo-displacement-controlled mode at 20 nm/s. The tests were terminated once either a load drop occurred or pillar splitting was visually observed during testing. The load-displacement behaviour was recorded and the maximum load at which splitting occurred was identified to calculate the pillar splitting fracture toughness. Since drift correction cannot be conducted during the pillar splitting tests, separate nanoindentation tests with a drift hold segment were performed immediately before and after the pillar splitting tests at each temperature to ensure temperature matching between the tip and sample and to estimate thermal drift rates. However, it is also important to note that for displacement-controlled testing thermal drift will mainly cause error in displacement measurements and not the load values [30]. During pillar splitting it is the critical load which is required for fracture toughness calculation, and hence thermal drift is not a critical factor here unlike in nanoindentation. *Post-mortem* microscopic examination was performed in Zeiss Gemini500 microscope to observe the fractured pillars.

*2.3 TEM investigation*

To characterise the deformation structure underneath the indents using TEM, cross-sectional specimens were prepared from the nanoindents using a Thermo Fisher Scientific Scios 2 FIB equipped with a Ga$^+$ ion source. In order to understand the effect of temperature on the deformation zone, TEM specimens were extracted from an indent made at room temperature (after cooling down from 580 °C) and an indent made at 450 °C, which lies in the BDT zone, and for which the indentation testing was stopped at this temperature without exposing the deformed microstructure to higher temperatures. TEM imaging and selected area electron diffraction (SAED) were performed on a JEOL JEM 2100+ operated at 200 kV and a C$_s$ image-corrected Thermo Fisher Scientific Titan Themis 60-300 equipped with the Ceta CMOS camera operated at 300 kV. To identify Burgers vectors of dislocations, TEM imaging was carried out in JEOL JEM F200 at 200kV under different two-beam conditions followed by *g.b* extinction analysis (where *g* is the diffraction vector and *b* is the Burgers vector).

## 3 Results
*3.1 BDT insights from high-temperature scratching*

Micro-scratch testing of the coarse-grained annealed microstructure provided initial insights into the temperature-dependent deformation behaviour of the *C*15 CaAl$_2$ intermetallic. Scratching was conducted from room temperature (RT) to 500 °C on grains oriented close to the (101) plane, as depicted in the EBSD inverse pole figure (IPF) map of **Fig. 1a**, with corresponding SE images of scratches at various temperatures shown in **Fig. 1b**. A scratch was also performed on the (001) surface plane at RT (marked as RT-2 in **Fig. 1b**), yielding results equivalent to those performed on the (101) surface. At RT for the (101) surface, minimal crack formation was observed at the scratch edges, although the presence of cracks was not always clear. The scratch width increased from ~350 °C, with crack formation less discernible at 400°C, accompanied by more uniform deformation, in particular at lower contact loads. The transition from 400 to 500 °C was more pronounced, revealing a smooth scratch profile without material pile-up. These variations in scratch deformation features imply a reduction in resistance to deformation at high temperature, resulting in wider and deeper scratches indicative of a more ductile deformation mode. However, due to the significant deformation variations observed over a wide temperature range from 350 to 500 °C, this method is unsuitable for accurately predicting the precise BDTT for CaAl$_2$.

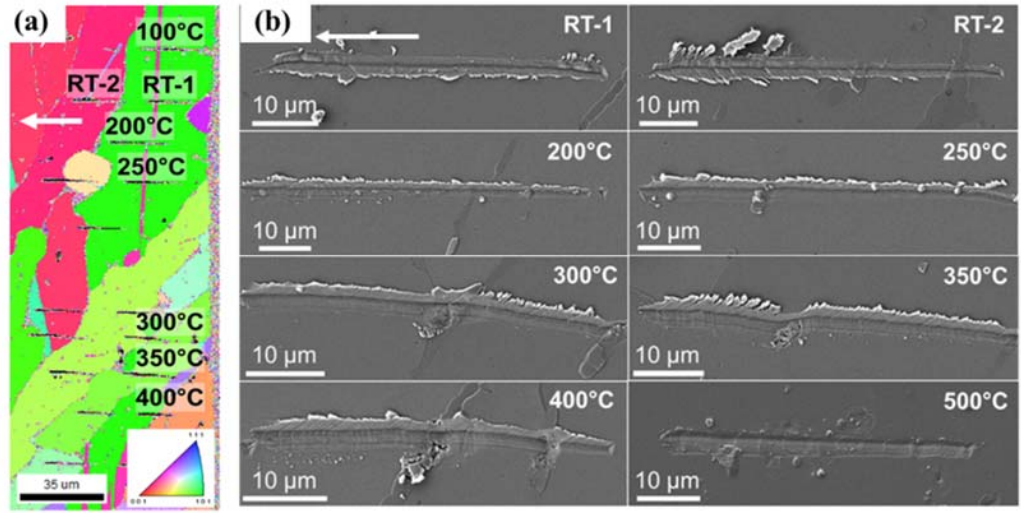

Fig. 1: *In situ* scratch testing from RT to 500 °C. (*a*) EBSD IPF map over the scratched surface highlighting the microstructure over which the scratches were performed; (b) secondary electron images of scratches after testing at different temperatures. The white arrow in the micrograph of RT-1 scratch indicates the direction in which the scratches were performed. The load was increased from 5 to 25 mN from the start (right side) to the end (left) of the scratches. All scratches at elevated temperature (100-500°C) were performed on the ~(101) surface, while at RT scratches were analysed for both the ~(101) (RT-1) and ~(001) (RT-2) surfaces.

*3.2 Effect of temperature on nanoindentation hardness*

Hardness variation with temperature was determined through nanoindentation tests to enhance understanding of how deformation behaviour changes with temperature in the *C*15 $CaAl_2$ phase. **Fig. 2a** shows the EBSD IPF map of the region where nanoindentation was performed, and the surface orientation of the tested grain has been marked. **Fig. 2b** shows the representative thermal drift-corrected load-displacement curves at temperatures from 25 to 580 °C. The load-displacement behaviour at 25 and 300 °C is also shown separately in the inset of **Fig. 2b**. With increasing temperature, clear differences are observed in the load-displacement behaviour in terms of: (i) progressive increase in indentation depth at the beginning of load-hold segment at the peak load from 500 °C; (ii) higher time-dependent displacement and maximum depth of indentation at the peak load from 450 °C. Both effects suggest that plastic deformability of $CaAl_2$ increases from ~450 °C, and is in broad agreement with the deformation changes observed from scratch testing in **Section 3.1**. This implies that nanoindentation hardness of the material varies with temperature, as observed in **Fig. 3**. The hardness at room temperature was ~4.5 GPa, similar to that recently reported for $CaAl_2$ [18]. The hardness remained constant at ~4.5 GPa up to 400 °C, and dropped by 22% at 450 °C to 3.5 GPa. This was followed by a more pronounced decrease at 500 °C to ~1 GPa. Subsequently, the hardness further reduced and was nearly constant at ~0.6 GPa between 550 to 580 °C. Thus, the temperature range between 450 to 500 °C marks the region where a significant enhancement in plastic deformation of $CaAl_2$ occurs in the cube corner nanoindentation tests. The observed time-dependent displacement during the hold segment (**Fig. 2b**) is attributed to creep deformation, influenced by high temperature and pressure in the centre of the plastic zone, as well as stress gradients from the cube corner geometry. This is distinguished from thermal drift, where displacement would vary linearly with time (discussed in **Section 4.3**). Comparison of hardness at different hold times (**Fig. 3**) reveals no significant difference at 500 °C. Longer hold times in the high-temperature regime slightly reduce hardness without altering the overall trend of saturation to lower values.

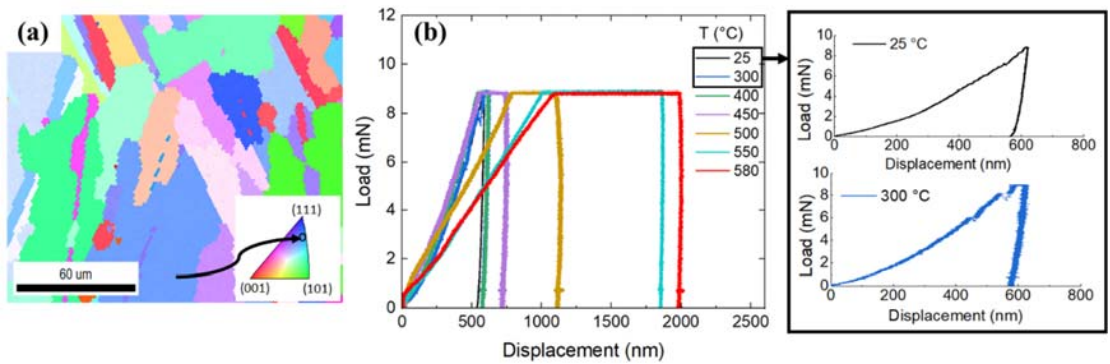

Fig. 2: High-temperature nanoindentation of $CaAl_2$. (a) Orientation map of $CaAl_2$ obtained from EBSD showing the grain orientation where nanoindentation tests were conducted. (b) Representative load-displacement curves at different temperatures obtained from nanoindentation; the inset in (b) shows separately the zoomed in version of load-displacement curves at 25 and 300 °C.

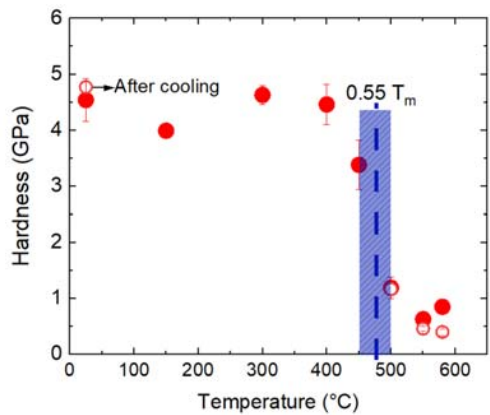

Fig. 3: Variation of nanoindentation hardness with temperature for the *C*15 CaAl$_2$ Laves phase. The filled circles denote the data for 10 s hold time and open circles are for 30 s hold time at peak load. Error bars indicate the standard deviation of hardness values at each temperature and for cases where it is invisible the error bar was smaller than the symbol size. The blue shaded region represents the transition temperature range where the hardness values decrease significantly from the upper plateau region before saturating to lower values.

Increased plastic deformation with temperature was also observed from the *post-mortem* SEM analysis of the residual indents. **Fig. 4** shows representative secondary electron images of the indents performed at various temperatures. The size of the residual indent impression is similar below 450 °C and then increases from 450 °C and above. A larger size of indent impression for fixed maximum load is consistent with the lower hardness at higher temperatures shown in **Fig. 3**. Notable changes in deformation features around the indent can also be observed with temperature. Slip steps and material pile-up are visible in the low temperature indents, along with cracks (indicated by white arrows) extending from the corners of the indents. No slip lines or cracks are visible in the transition and high temperature regime above 300 °C. Pile-up diminishes noticeably at 450 °C and is not observed at ~500 °C. Surface features appear less defined in post-cooling images due to surface oxidation at elevated temperatures. It must be emphasised that even with high vacuum level ($< 5 \times 10^{-6}\ mbar$) in the SEM, surface oxidation of CaAl$_2$ cannot be avoided (see presumed oxide scale of ~60 nm thickness observed from the TEM cross-section of an indent made at RT after exposure to 580 °C in **Fig. 9a**). These changes in deformation features with temperature, in conjunction with the significant decrease in hardness, suggests that the deformation behaviour of the CaAl$_2$ Laves phase changes at ~450 °C and is facilitated by higher plastic deformability with the absence of cracking. To further correlate the changes in hardness with changes in fracture behaviour across this transition zone, high-temperature micropillar splitting tests were conducted.

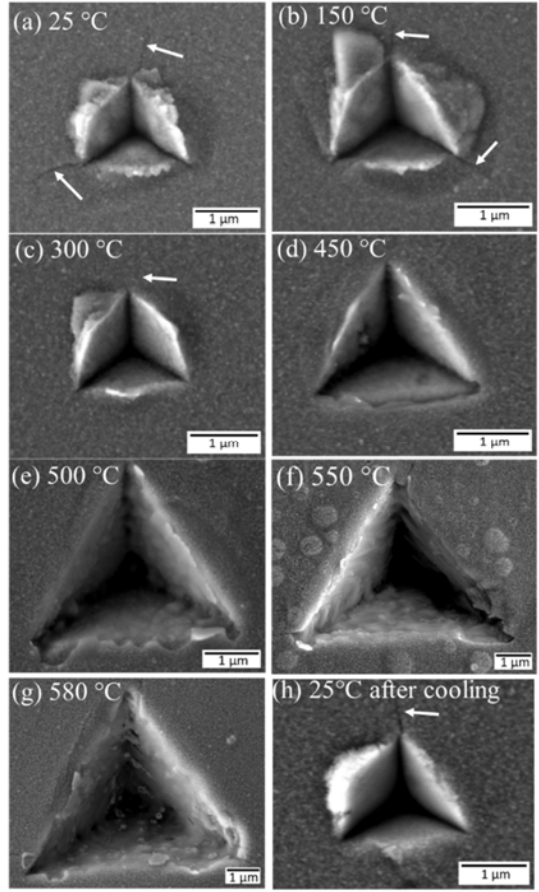

Fig. 4: Secondary electron *post-mortem* images of nanoindents made at various temperatures showing changes in slip lines and crack formation (marked with white arrows) around indents. From 25 to 450 °C all images were taken at the same magnification, whereas from 500 to 580 °C a lower magnification was used to capture the larger indents.

### 3.3  *Effect of temperature on the pillar splitting fracture toughness*

**Fig. 5a** shows an overview of micropillars fabricated on a single CaAl$_2$ grain, whose surface orientation is marked in the inset IPF. A representative high-magnification image of a representative micropillar is also shown in the inset in **Fig. 5a**. **Fig. 5b** shows a micrograph of the *in situ* configuration with cube corner indenter prior to testing, while **Fig. 5c** shows the representative load-displacement response of the pillar splitting tests at various temperatures. The load-displacement curves maintain a consistent shape up to 300 °C, showing minimal variation in slope and indicating negligible effects from indenter positioning and thermal drift. The maximum load at which pillar splitting occurs (identified by a sharp load drop) increases only slightly from 25 to 150 °C, and then increases significantly at 300 °C. A transition is observed at 400 °C, wherein the peak load increases slightly, but pillar splitting does not occur.

Above 400 °C, the peak load at the pre-set maximum indentation depth decreased, signifying that the CaAl$_2$ intermetallic becomes more plastically deformable. Due to lack of pillar splitting at 400 and 500 °C, a load drop was not observed up to the maximum set displacement value of 2.5 μm, where the tests were terminated. Hence, the apparent fracture toughness was determined only up to 300 °C. The critical stress intensity factor for splitting was evaluated using **Eq. 2** [27]:

$$K_c = \gamma \frac{P_c}{R^{1.5}} \tag{2}$$

which assumes an isotropic material volume. $K_c$ is the apparent fracture toughness, $P_c$ the critical load at which pillars split, $R$ is the radius of the pillar, and $\gamma$ is a factor which depends on the geometry of the indenter and $E/H$ ratio of the material. A constant elastic modulus $E$ of 140 ± 6 GPa was used between 25 and 300 °C, based on modulus values obtained from unloading segment of the nanoindentation experiments. It is to be noted that no positive displacement due to creep occurred during unloading up to 300 °C. Using the hardness values at each temperature obtained from nanoindentation measurements in **Section 3.1**, the $E/H$ ratio was obtained within the constant $\gamma$ range for a cube corner indenter geometry reported from Ref. [42] (i.e. ≈31), and therefore $\gamma \approx 0.85$ was applied over the studied temperature range of 25 to 300 °C. **Fig. 5d** shows the variation of apparent fracture toughness from pillar splitting as a function of temperature along with the maximum load $P_{max}$ for all temperatures. $P_{max}$ corresponds to $P_c$ for the case where pillar splitting occurred from 25 to 300 °C, whereas at 400 and 500 °C it was the maximum load till the displacement set-point was reached. The $K_c$ calculated from **Eq. 2** was 0.94 ± 0.05 $MPa\sqrt{m}$ at 25 °C, and increased slightly to 1.2 $MPa\sqrt{m}$ at 150 °C, whereas at 300 °C it increased significantly to 2.76 ± 0.26 $MPa\sqrt{m}$ with a commensurate increase in the critical load for splitting. Pillar splitting was observed again after cooling back to 25 °C with $K_c$ ~0.9 ± 0.1 $MPa\sqrt{m}$, implying that exposure to high temperature did not induce any major structural changes in the material to influence the toughness, and that surface oxidation did not severely impact the toughness measurement, similar to the nanoindentation results in **Section 3.2**.

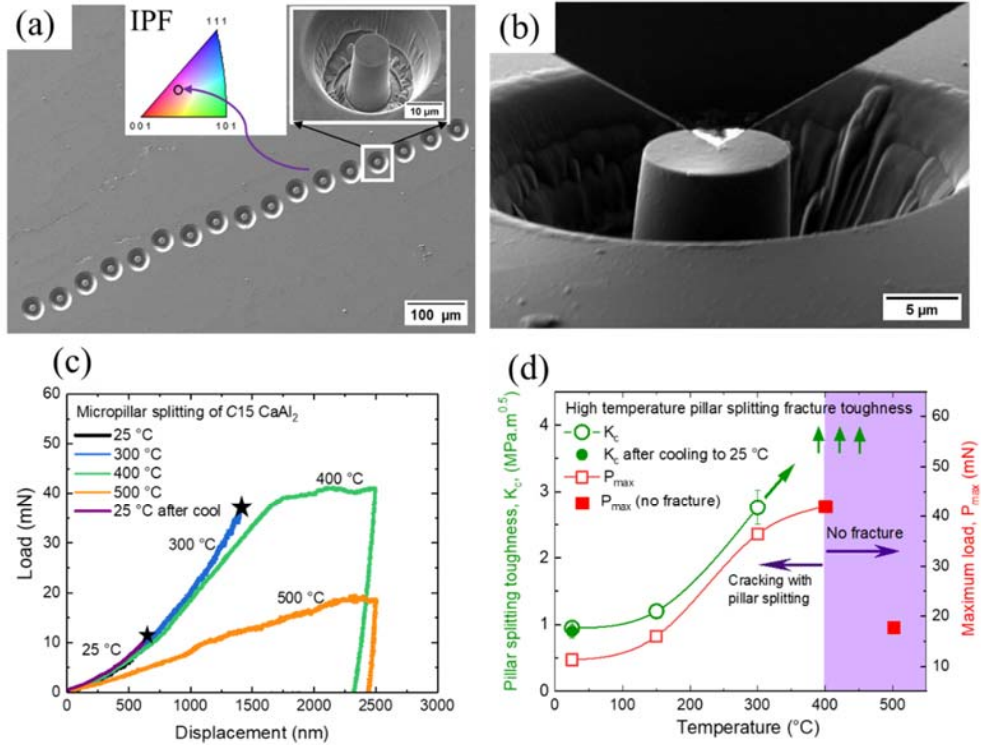

Fig. 5: (a) Secondary electron overview image of micropillars fabricated in a single grain whose surface orientation is indicated in the IPF (inset) together with a high magnification tilt-corrected image of a single micropillar at 45° tilt angle; (b) Micropillar loaded with a cube corner indenter prior to testing at 300 °C; (c) Representative load-displacement curves for pillar-splitting at various temperatures; (d) Variation of pillar-splitting fracture toughness and maximum load $P_{max}$ with temperature. The filled red squares in (d) indicate the maximum load for the cases where no splitting occurred. The green arrows in (d) indicate that apparent fracture toughness would increase with temperature above 300 °C.

**Fig. 6** shows *post-mortem* images of the pillars after the splitting tests at different temperatures. From 25 to 300 °C straight cracks predominantly extend from the corner of the indents, leading to 3-way splitting of the pillars. In few cases, due to inaccuracy in indenter positioning, an ideal 3-way splitting did not occur wherein two of the cracks extending from the indenter corner reached the edge of the pillar while the third did not. At 400 and 500 °C no critical crack initiation for pillar splitting was observed, although some isolated tortuous cracks are present on the surface of the deformed pillars. Such surface cracks could be due to stresses arising from accommodation of the larger plastic zone below the indent, or cracking of the surface oxide layer. However, these surface cracks do not lead to pillar splitting. Crack formation was not observed during nanoindentation at similar temperatures due to the hydrostatic constraint from the surrounding material, which is reduced in the case of a micropillar volume. For the lower temperatures where pillar splitting did occur, the crack planes were indexed using the method used for surface trace analysis discussed in Ref. [18]. Some of the crack planes have been marked in **Fig. 6** and belong to low index planes. However, it is not possible to discern any possible change in crack plane with temperature in this case, as cracking would primarily be driven by the cleavage planes which are aligned with the sharp corners of the indenter. In our earlier study [19], the fracture toughness values of different low index cleavage planes in $C15$ $CaAl_2$ were found to be similar. Hence the orientation of the grain and alignment of the cleavage planes with indenter diagonal will likely not significantly impact the pillar splitting fracture toughness value of $C15$ $CaAl_2$.

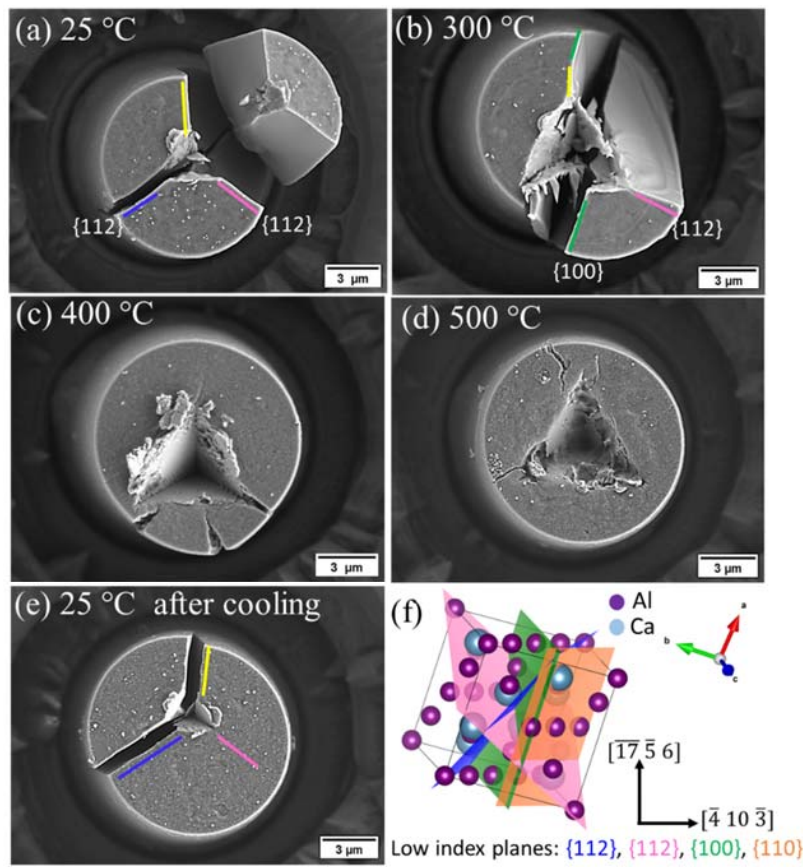

Fig. 6: *Post-mortem* secondary electron micrographs, showing the top view after pillar splitting tests at (a) 25 °C, (b) 300 °C, (c) 400 °C, (d) 500 °C and (e) after cooling back to 25 °C. (f) Crystal orientation of the grain showing potential low-index planes for cracking; the in-plane orientations of pillars in (a-e) are indicated in (f). The yellow lines in (a), (b) and (e) indicate crack planes for which two possible solutions were obtained and hence could not be assigned unambiguously.

## 4   Discussion

### 4.1   BDTT of C15 CaAl$_2$ Laves phase: A comparison of the test methods

The results presented in **Section 3** highlight a clear change in the deformation behaviour of *C*15 CaAl$_2$ Laves phase with increasing temperature, and is supported by three distinct changes: (i) significant drop in nanoindentation hardness at 450-500 °C; (ii) absence of cracking at the corners of indents at 300 °C and above; (iii) brittle pillar splitting between 25 to 300 °C, followed by no splitting from 400 °C. These changes imply an overall transition from brittle to a more ductile behaviour of CaAl$_2$, accompanied by a higher contribution of plasticity to fracture. Disappearance of cracks at ~300 °C during cube corner nanoindentation suggest that crack tip shielding due to local dislocation activity is already operative at these lower temperatures, which together with the hydrostatic constraint from the surrounding material may limit crack extension. This is consistent with the increase in pillar splitting fracture toughness at 300 °C. During micropillar splitting, unstable crack initiation and extension occurs at the critical load for instability. Enhanced dislocation activity under the indenter tip at higher temperatures can better accommodate local stress thereby affecting the critical load for crack initiation during the splitting process. However, brittle splitting fracture of the micropillars is still observed at 300 °C, implying that CaAl$_2$ remains brittle at this temperature.

The influence of local plasticity is more evident as temperature increases to 400 °C, where the maximum load reaches a peak value (see **Fig. 5**), however no splitting occurs, indicating that CaAl$_2$ becomes less brittle at this temperature. Ast *et al.* [43] reported that for crack nucleation and pillar splitting to occur within a micropillar volume size which can be typically achieved by FIB milling, the $K_c/H$ ratio should be less than 0.8 for a cube corner indenter. For our system at 300 °C the ratio is 0.61, whereas at 400 °C it would increase to 0.88 if one assumes that the $K_c$ increases with temperature following a linear extrapolation beyond 300 °C (see green arrow in **Fig. 5d**). Since pillar splitting is limited to materials which fail in a brittle manner, it is unsuitable for obtaining a high-temperature toughness for which alternate elastic-plastic fracture mechanics approaches are required. Interestingly, while local dislocation activity can blunt the crack tip, increase fracture toughness and even prevent pillar splitting at 400 °C, no decrease in indentation hardness or increase of the residual indent size was observed (see **Figs. 3** and **4**). Only at and above ~450 °C does the hardness drop and dislocation plasticity becomes more dominant (see TEM analysis discussed in Section 4.2).

The transition zones between brittle and ductile behaviour of CaAl$_2$ are shown in **Fig. 8,** where both the temperature-dependent variation of hardness and pillar splitting toughness have been compared. Up to 300 °C, CaAl$_2$ exhibits brittle behaviour with constant hardness and brittle failure due to cracking. Between 300 and 400 °C a microplasticity zone is present where although the hardness does not change, cracks at indenter edges are not observed during nanoindentation, while significant local dislocation activity at the indenter apex and possible crack tip shielding effects increase the toughness to an extent that a critical fracture initiation could not be determined from micropillar splitting tests. The microplastic region above 300 °C is also supported by the morphological changes in wear tracks from scratch testing (**Fig. 1**) as this method primarily probes the surface wear deformation. At approximately 450 °C and above, CaAl$_2$ exhibits enhanced plasticity due to dislocation activity, as evident from the significant drop in hardness together with absence of pillar splitting, thereby transitioning into a more ductile behaviour. The close correspondence in the temperatures where plasticity prevents the brittle failure of pillars (~400 °C) and decreases nanoindentation hardness (~ 450 °C) suggests that a prominent change over into the ductile regime will occur from 450 °C (~0.55 $T_m$).

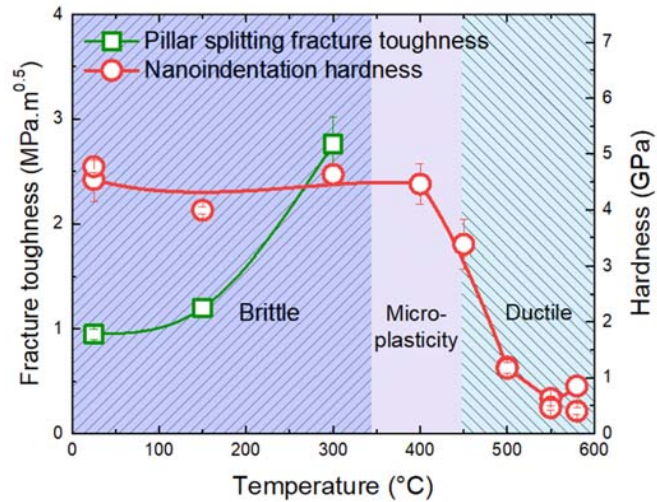

Fig. 8: Comparison of nanoindentation hardness and pillar splitting fracture toughness with temperature, highlighting brittle and ductile deformation zones.

The nanoscale BDTT (~0.55 $T_m$) of *C*15 CaAl$_2$ determined here is lower than the macroscale BDTT of ~0.65 $T_m$ typically reported for other *C*15 Laves phases via macro-compression testing [4], where BDTT was defined as the temperature at which a compressive strain of 1-2 % could be achieved without specimen fracture. However, from Vickers indentation of *C*15 NbCr$_2$ a BDTT of ~0.55 $T_m$ was obtained [44], equal to that obtained for *C*15 CaAl$_2$ in this study. Further, to contextualise the BDT behaviour of CaAl$_2$ determined here using small-scale testing methods in relation to macroscale testing, the microplasticity zone at small length-scales suggests a wider temperature range for complete transition into the ductile regime. In comparison, during macroscale tests a more abrupt BDTT could occur for intermetallic compounds [9,45,46].

Method and length-scale-dependent BDTT differences are evident in other literature examples, where various testing methods were employed. The toughness reported from Si micropillar splitting increased at ~175 °C due to local crack tip shielding possibly facilitated by dissociation of dislocations into partials [30,47], whereas the microhardness decreased at higher temperatures of 500 °C due to enhanced dislocation activity [48]. Additionally, high-temperature microcantilever fracture tests on Si demonstrated a progressive increase in mode I fracture toughness from ~300 °C to 600 °C [31]. On the other hand, macroscale testing using pre-cleaved Si single crystals revealed a sharp increase in fracture toughness at 705 °C [49]. This supports our observation that small-scale toughness increases at lower temperatures when compared to the indentation hardness. Similar length-scale effects have been observed for tungsten [32], where the BDTT obtained from macroscale testing of three- and four-point bending specimen geometries was ~150 to 200 °C [50,51], while microcantilever fracture tests revealed a BDTT of ~0 °C [32], with a gradual increase in fracture toughness and change from brittle-cleavage to micro-cleavage fracture due to increasing dislocation activity. These studies suggest that the onset of BDTT, facilitated by dislocation activity enhancing fracture toughness and promoting ductile behaviour, may occur at lower temperatures on small length scales. Differences based on the testing method, evident in case of CaAl$_2$, can also arise.

Regarding surface oxidation challenges for CaAl$_2$ at high temperatures, for micro-scratching experiments the formation of an oxide scale obscures slip trace observation near the wear track and quantitative insights into elastic and plastic wear depth or friction coefficient may become unreliable [36]. Additionally, it must be noted that due to the observed oxidation, BDTT determination using notched microcantilever tests is not possible to the changing stress/chemistry at the notch root, which can strongly affect the measured fracture toughness [52]. Moreover, the plastic zone size $r_p$ ahead of the crack tip of notched microcantilevers can increase with temperature and exceed the cantilevers dimensions, leading to size effects on cantilever fracture toughness. As such, microcantilever fracture tests were not performed here, and our micro-scratch tests were interpreted with a degree of caution.

*4.2 Mechanistic insights into the thermally activated deformation of C15 CaAl₂*

The deformed microstructure beneath the indents made at RT and at the BDTT, were analysed to examine the effect of temperature on dislocation structure. Site-specific lamellas were extracted from below two representative indents made at 25 °C and 450 °C and examined by TEM. A schematic representation of the orientation in which the TEM lamella was extracted from indents has been given in **Fig. S2** of the Supplementary Information. **Figs. 9a,b** show the bright-field TEM (BF-TEM) and BF-scanning TEM (BF-STEM) images of the deformed region underneath the indent at 25 °C, respectively. Straight parallel slip bands are visible, as indicated by the solid red arrows, along with dislocation entanglement in the plastic zone of the indent at 25 °C (refer to **Fig. S3** for TEM images obtained from weak beam imaging by tilting to different two beam conditions). Similar slip bands in the deformed microstructure below nanoindents in the same $C$15 CaAl₂ sample were observed by Freund *et al.* [18] for Berkovich indents at RT. Furthermore, a crack is visible at the bottom of the indent (indicated by a dashed yellow arrow), which is consistent with SEM observation of surface cracks around the RT indent shown in **Fig. 4**. **Fig. 9c** shows the BF-TEM image of the plastic zone below a nanoindent at 450 °C. In contrast to the RT indent, no straight slip lines or cracks are observed. Rather, a high density of homogenously distributed dislocations, forming a network of curved dislocation lines, is present in the vicinity of the indent, with some isolated dislocations extending to farther regions. The absence of slip lines and cracks supports the SEM observations of homogeneous deformation without formation of slip traces on the indent surface at 450 °C in **Fig. 4d**, and the lower hardness due to more uniform dislocation activity. It is noted that due to the load-controlled nanoindentation tests performed here, the deformation zone of the indents at 25 and 450 °C were not of equivalent maximum depth. It is reported for the same $C$15 CaAl₂ sample used in this study that $a/2<110>$ (where $a$ is the crystal lattice parameter) type perfect dislocations, as well as stacking faults on {111} planes are present within the slip bands at RT [18]. The latter were thought to stem from the motion of $a/6<112>$ dislocations. The TEM images of **Fig. 10** show that in case of the high temperature indent lamella (of **Fig. 9c**), tilting to different zone axes and two-beam conditions also revealed similar $a/2<110>$ perfect dislocations (indicated with solid white arrows) and partial dislocations of $a/6<112>$ type (indicated with dashed arrows). These dislocations were identified by performing a $g \cdot b$ analysis on visible/invisible dislocation sets obtained from different two-beam conditions. A set of visible dislocations for zone axis [101] (in **Fig. 10a**) and g= ($1\bar{3}\bar{1}$) are shown in **Fig. 10b**, while the dislocations become invisible with g=($22\bar{2}$) as shown in **Fig. 10c**. Refer to **Fig. S4** in the Supplementary Information for further TEM images of dislocation structure below this indent for other two beam conditions. Although resolving the partial dislocations in the RT sample may simply be impeded by the overlapping strain fields where they are concentrated in planes of high defect density, the observation of partial dislocations at elevated temperature is consistent with synchroshear requiring substantial thermal activation to operate [54] and previous TEM investigations on the structurally related Nb-Co μ-phase, which also exhibits a transition from full to partial dislocation slip [55] within the Laves phase building block of its larger unit cell, also associated with a change in slip plane that cannot be resolved by conventional TEM [56].

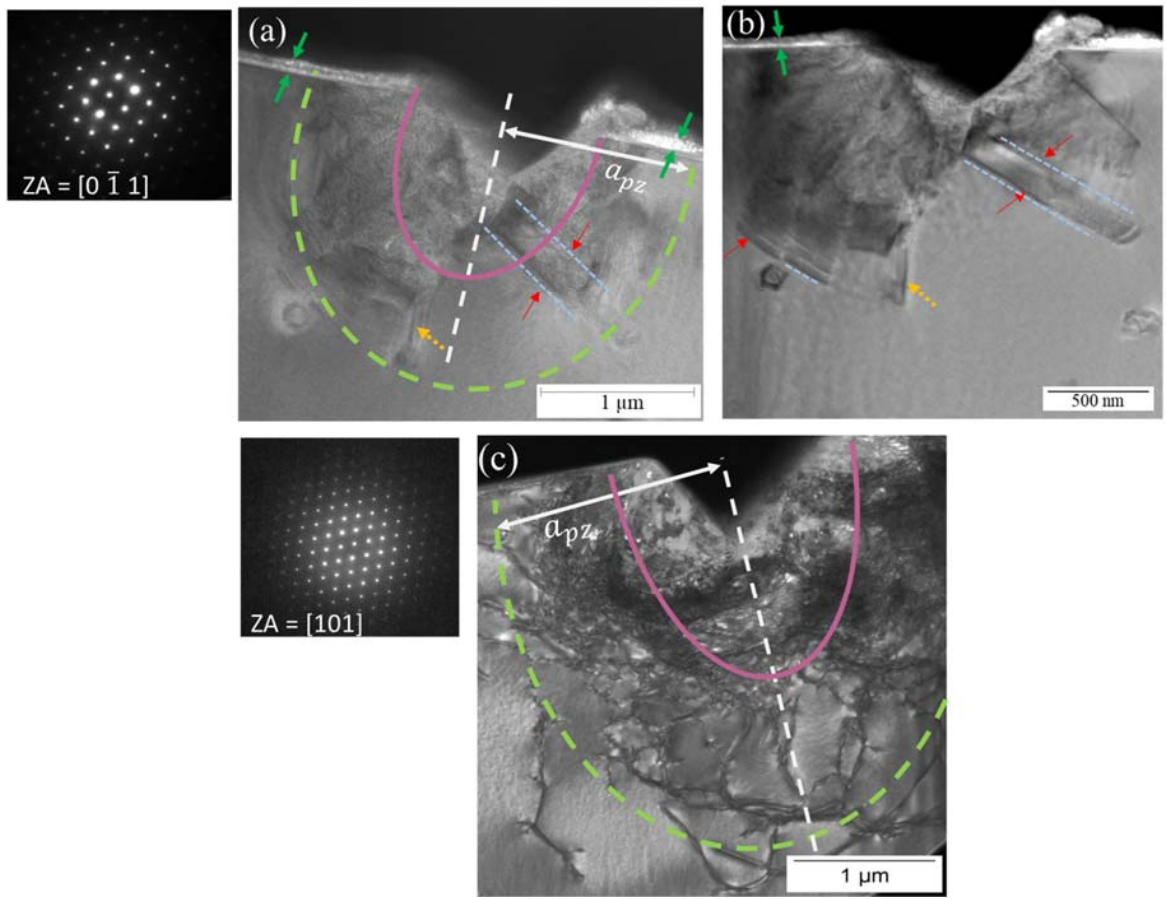

Fig. 9: TEM images showing the cross-section of the deformed region under nanoindents performed at different temperatures. (a) BF-TEM and (b) BF-STEM images of indent at 25 °C taken in the [$0\bar{1}1$] zone axis. The orientation of the specimen is seen by the SAED pattern shown. The red arrows in (a) and (b) show the activated slip planes and the yellow arrows mark the crack observed at 25°C; the green

arrows in (a) and (b) indicate presumed oxide layer formed on the surface of indent after exposure to 580 °C. (c) The BF-TEM image of indent at 450 °C shows no slip plane activation or crack formation and the SAED pattern shows that the BF-TEM image was acquired at the [101] zone axis. In (a) and (c) the theoretically computed plastic zone has been shown by the solid magenta curve for $f$=1 and green dashed curve for $f$=2.4 (see main text for more details). The dark region above the indented surface in (a-c) is the Pt layer.

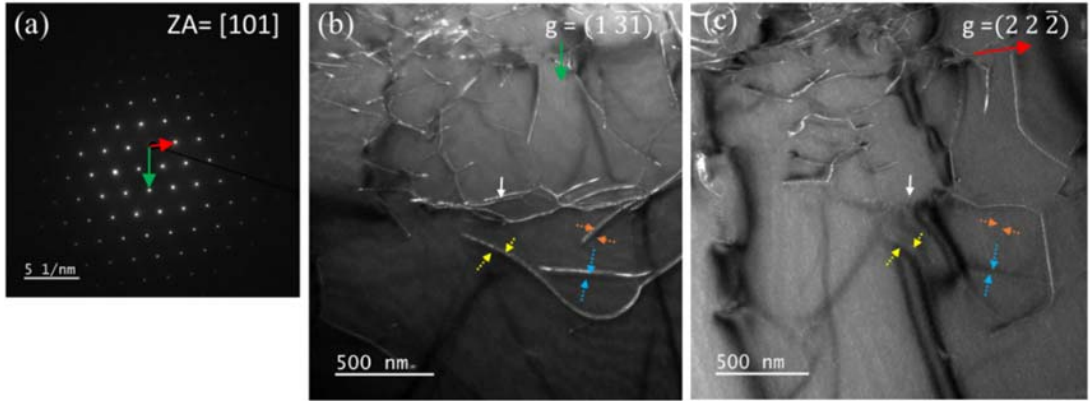

Fig. 10: Dislocation structure under different two-beam conditions underneath the nanoindent at 450 °C. (a) Diffraction pattern of [101] zone axis, and weak-beam dark field images under (b) g=(1$\bar{3}$1) where the dislocations are visible and (c) g=(22$\bar{2}$) where the dislocations are invisible. Solid arrows indicate perfect dislocation whereas dashed arrows of same colours denote partial dislocations identified from the *g.b* analysis.

**Fig. 9** highlights that the microstructural features within the deformed region changes with temperature, accompanied by an increase in plastic zone size. To compare between the observed plastically deformed region below the indents with the theoretical prediction, the plastic zone size $a_{pz}$ was calculated according to the modified Nix-Gao model [57,58], where $a_{pz} = f a_c$ ($a_c = h/\tan\theta$, $\theta$ is the angle between surface and indenter and $h$ is the indentation depth). The factor *f* (equal to 1 in the original Nix-Gao model [59], and greater than 1 in the modified one [57,58]), relates the contact radius to the plastic zone size, and depends on the material and its crystallographic orientation. From finite element simulations of cube corner indentation of FCC Cu in Ref. [58], $f$=2.4 was estimated. The theoretically predicted plastic zone regions, according to $a_{pz} = f a_c$, are indicated for both temperatures in **Figs. 9a,c**, for both $f$=1 and $f$=2.4 and with $\theta = 47.7°$ for the cube corner indenter. It is observed that for $f$=2.4 (where $a_{pz}$ = 1280 nm at 25 °C and 1580 nm at 450 °C), the theoretically predicted plastic zone size matches reasonably with the deformed zones observed from TEM, although it appears slightly larger for the indent at 25 °C. It is to be noted that some differences can occur due to slight misalignment of the lamella with respect to the indent centre, the effect of stresses at the corner of the indent and also due to the tendency of the intermetallic phase to highly concentrate deformation onto individual planes at ambient temperature. This increase in the plastic zone size with depth of indentation of $CaAl_2$ Laves phase, equivalent to the behaviour of FCC metals, confirms that the observed decrease in hardness and the BDT of $CaAl_2$ intermetallic is controlled by enhanced dislocation mediated plasticity at elevated temperatures which can blunt the cracks and restrict brittle failure.

The limited plasticity of Laves phases below the BDTT is generally attributed to restricted dislocation activity within their complex close-packed structures. Initially, synchroshear was proposed as the sole dislocation mechanism [60–62]. A high lattice friction for dislocation glide has also been considered to hinder dislocation activity at RT for Laves phases [63,64]. From the room temperature activation volume of $C$15 $CaAl_2$ (~$3.75b^3$, where $b$ is the Burgers vector) [18], it was not possible to definitively determine if the deformation was governed by overcoming Peierls resistance. Here, the RT lattice friction stress of $CaAl_2$ intermetallic was estimated from the nanoindentation hardness data in **Section 3.2** by following the method proposed by Qui *et al.* [65], as outlined in the Supplementary Information. Considering a perfect $a/2<110>$ dislocation with $b=0.5656$ nm for this analysis in $CaAl_2$, the friction stress is estimated as 580 MPa, which lies in the range of values of friction stress (0.4 to 5.6 GPa) reported for other $C$15 Laves phases [66].

The enhancement of plasticity above the BDTT for many transition metal $C$15 Laves phases has been attributed to increased dislocation motion due to thermally activated processes such as creep deformation [4], while for some Laves phases it has been attributed to a Peierls mechanism due to high strain rate dependence and low activation volume [25,67]. Noticeable creep deformation in $CaAl_2$ can be observed in **Fig. 2b** from the increased time-dependent displacement during the constant load hold segment in the nanoindentation tests at high temperatures (450 °C and above). The BDTT of the $CaAl_2$ Laves phase (~0.55 $T_m$) in the present case as well as other transition metal $C$15 Laves phases [4], is above a homologous temperature of 0.5 $T_m$ consistent with temperatures required for thermally-activated creep process. The indentation creep behaviour of $CaAl_2$ analysed from the indentation strain rate and hardness obtained from nanoindentation results is presented in **Fig. S5** of the Supplementary Information. Interestingly, the indentation stress exponent of ca. ~3 to 3.2 obtained for CaAl$_2$ is similar to the creep stress exponent of approximately 3.5 reported from compression creep tests of other *C*15 Laves phases [4], and is indicative of dislocation creep [68]. Decoupling additional effects of temperature-dependent critical resolved shear stress and effect on high temperature plasticity would require further studies using micropillar compression. Overall, the high BDTT of 450-500 °C obtained for CaAl$_2$ suggests that this intermetallic will not undergo a loss in its strength due to enhanced plasticity at the elevated temperatures of ~200 °C desired for the Mg rich alloys [12], thereby maintaining the overall alloy strength.

## 5    Conclusions

- Indentation-based testing using nanoindentation and micropillar splitting as well as scratch testing were used to estimate the nanoscale BDT of the *C*15 CaAl$_2$ Laves phase. Scratch testing revealed plasticity at and above 350 °C, but was prone to surface oxidation and thus difficult to reliably analyse. In contrast, nanoindentation and micropillar splitting provide quantitative insights less prone to surface oxidation.
- A significant decrease in nanoindentation hardness occurred at ~450-500 °C, with distinct changes in the deformation features around the indents due to increase in size of the residual impression at and above 450 °C and disappearance of surface cracks around 300 °C. This implies a change in the deformation behaviour with temperature for the CaAl$_2$ Laves phase.
- Elevated temperature micropillar splitting revealed brittle pillar splitting up to 300 °C, with straight crack propagation from the centre to the edge of the pillar. The apparent fracture toughness remained nearly constant up to 150 °C, followed by a sharp increase at 300 °C. No pillar splitting occurred at 400 °C and above due to considerable plastic deformation.
- A comparison of the scratch tests, nanoindentation and micropillar splitting results indicate that CaAl$_2$ remains brittle up to 300 °C, followed by increasing microplasticity between 300-400 °C which can cause crack tip shielding but not affect the nanoindentation hardness. Finally, at ~450 °C and above a transition to a more ductile behaviour occurs with a pronounced decrease in hardness due to a more uniform dislocation activity and the absence of brittle pillar splitting.
- TEM examination of the deformed microstructure below the indents revealed slip bands and small cracks in the deformed zone under an indent made at 25 °C. At 450 °C, the deformation zone extends ~25% further and reveals a more uniformly distributed entanglement of dislocations.
- Nanoindentation hold segments reveal creep with creep rate exponents indicative of dislocation creep at temperatures above the BDTT, suggesting that the BDT of $CaAl_2$ can be driven by thermally-activated effects on dislocation motion leading to creep.

**Declaration of Conflict of Interest**

The authors declare they have no conflict of financial interest or personal relationships that could have influenced the results reported in this study.

**Acknowledgement**

The authors gratefully acknowledge the financial assistance from Deutsche Forschungsgemeinschaft (DFG) within projects B06, A05 and Z of Collaborative Research Center (SFB) 1394 "Structural and Chemical Atomic Complexity - from defect phase diagrams to material properties", project number 409476157. Authors thank Uzair Rehman for facilitating the high temperature scratch experiments, and Leon Christiansen and Dr. Sebastian Schädler (Carl Zeiss Microscopy Customer Center, Europe) for assistance in micropillar fabrication. The authors also thank Dr. Frank Stein for meaningful scientific discussions.

*Supplementary Material for*

# Nanoscale brittle-to-ductile transition of the *C*15 CaAl$_2$ Laves phase

Anwesha Kanjilal[1], Ali Ahmadian[1,a)], Martina Freund[2], Pei-Ling Sun[2], Sandra Korte-Kerzel[2], Gerhard Dehm[1,*] and James P. Best[1,*]

[1] *Max-Planck-Institut für Eisenforschung GmbH, 40237 Düsseldorf, Germany*
[2] *Institut für Metallkunde und Materialphysik, RWTH Aachen University, 52056 Aachen, Germany*
[a)] *Present address: Karlsruhe Institute of Technology, Institute of Nanotechnology, 76433 Eggenstein-Leopoldshafen, Germany*
[*] *Corresponding authors: j.best@mpie.de and dehm@mpie.de*

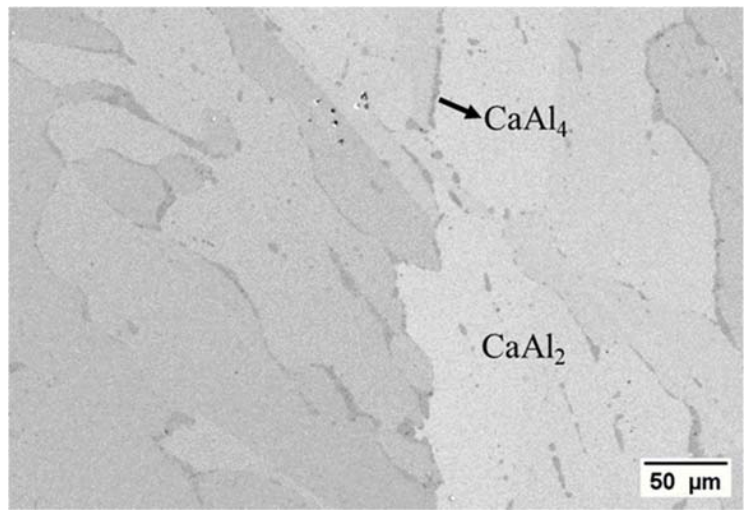

**Figure S1**: Representative secondary electron image of single-phase cast and annealed CaAl$_2$, with a skeleton network of CaI$_4$ intermetallic formed along the grain boundaries.

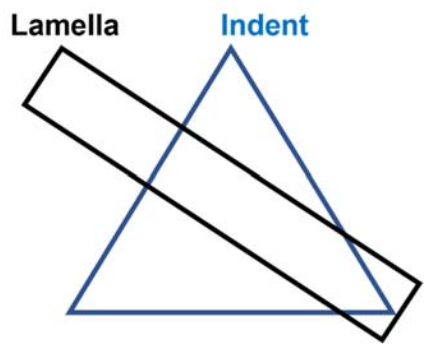

**Figure S2**: Schematic representation of the orientation in which TEM lamellae were extracted from the nanoindents.

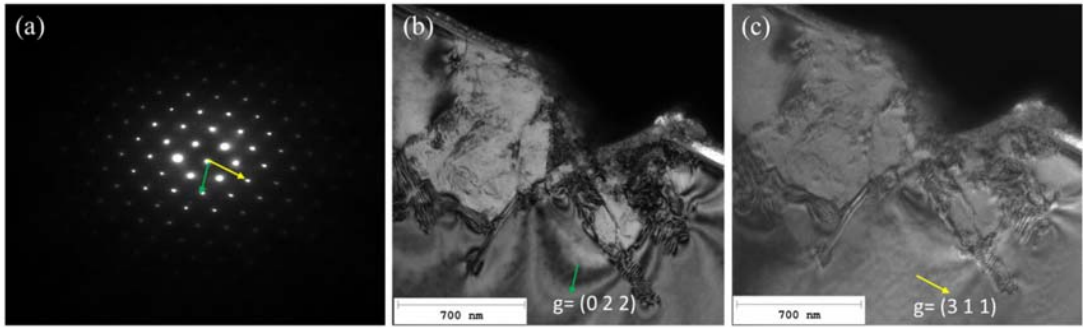

**Figure S3**: Dark-field TEM images of dislocation structure below the nanoindent at 25 °C shown in **Fig. 9b**, taken under different two-beam conditions. (a) Diffraction pattern of [0$\bar{1}$1] zone axis, and TEM images taken in [0$\bar{1}$1] zone axis with (b) $g_{022}$ and (c) $g_{311}$ diffraction contrast.

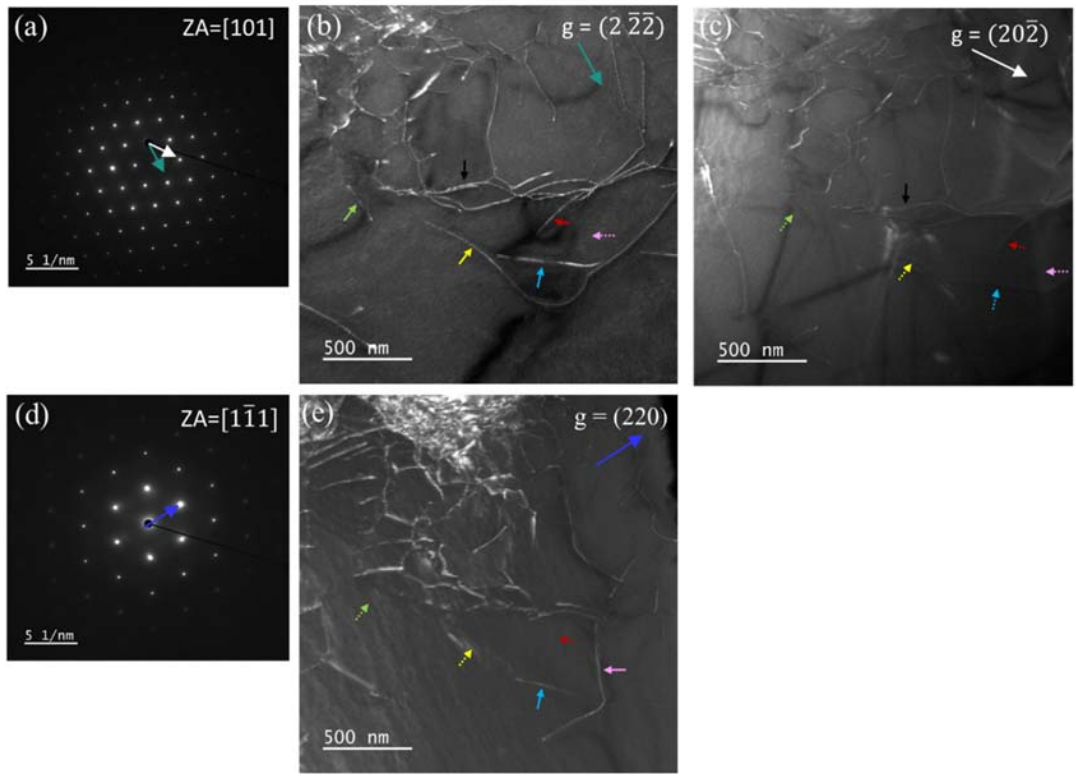

**Figure S4**: Dark-field TEM images of dislocation structure below the nanoindent at 450 °C shown in **Fig. 10**, taken under the zone axes (a) [101] and (d) [1$\bar{1}$1]. The corresponding dislocation structures are shown for different two beam conditions with (b) g=(2$\bar{2}\bar{2}$) and (c) g=(20$\bar{2}$) for [101] zone axis and (e) g=(220) for [1$\bar{1}$1] zone axis. The visible dislocations are marked with solid arrows and the equivalent invisible dislocations are shown with dashed arrows.

**Estimation of lattice friction stress of CaAl₂ at room temperature**

The Taylor dislocation model [1] relates the shear stress, $\tau$, with the dislocation density, $\rho$, as follows:

$$\tau = \tau_o + \alpha G b \sqrt{\rho_{SSD} + \rho_{GND}} \qquad (S1)$$

where $G$ is the shear modulus, $b$ is the dislocation Burgers vector, $\rho_{SSD}$ is the statistically stored dislocation density for a representative strain and $\rho_{GND}$ is the geometrically necessary dislocation density. It is assumed that the von Mises flow rule is valid with $\sigma = \sqrt{3}\tau$. The friction stress $\tau_o$ due to intrinsic lattice resistance was determined from the nanoindentation hardness $H$ using the method proposed by Qiu *et al*. [2]. For materials with high lattice friction, the total contribution of hardening due to friction stress and due to dislocations to the measured hardness can be given as follows [2]:

$$H = c\sigma = 3\sigma = 3\sigma_o + 3M\alpha G b \sqrt{\rho_{SSD} + \rho_{GND}} \qquad (S2)$$

where $\sigma_o = M\tau_o$, $\alpha$ is a constant taken as 0.5. $M$ is the factor to convert equivalent flow stress to hardness and is equal to $\sqrt{3}$ assuming that the von Mises flow rule applies. It is noted that for the sake of simplicity the constraint factor $c$ was taken as 3 in Eq. S2, although for Laves phases it can be lower. The GND density can be computed based on Durst *et al*. [3] as follows:

$$\rho_{GND} = \frac{3 \tan^2 \theta}{2 f^3 b h} \qquad (S3)$$

where $\theta$ is the angle between surface and indenter, $f$ is a factor which relates the contact radius $a_c$ to the radius of plastic zone $a_{PZ}$, and depends on the type of material and its crystallography, and $h$ is the depth of indentation. For cube corner indenter, assuming constant $f=2.4$ [4], and $\theta = 47.7°$, the GND density for CaAl₂ can be estimated from Eq. S3. The statistically stored dislocation density, $\rho_{SSD}$, depends on the macro-hardness as follows [2]:

$$H_{macro} = 3\sigma_o + 3\sqrt{3}\alpha G b \sqrt{\rho_{SSD}} \qquad (S4)$$

where $H_{macro}$ is the room temperature macro-hardness computed from the uniaxial stress-strain response. Due to lack of uniaxial data for the brittle CaAl₂ intermetallics, it is assumed that $H_{macro}$ is equivalent to Vickers hardness of CaAl₂ [5]. The lattice friction stress can then be estimated using equations S2 and S4.

# Indentation creep behavior of CaAl$_2$ Laves phase

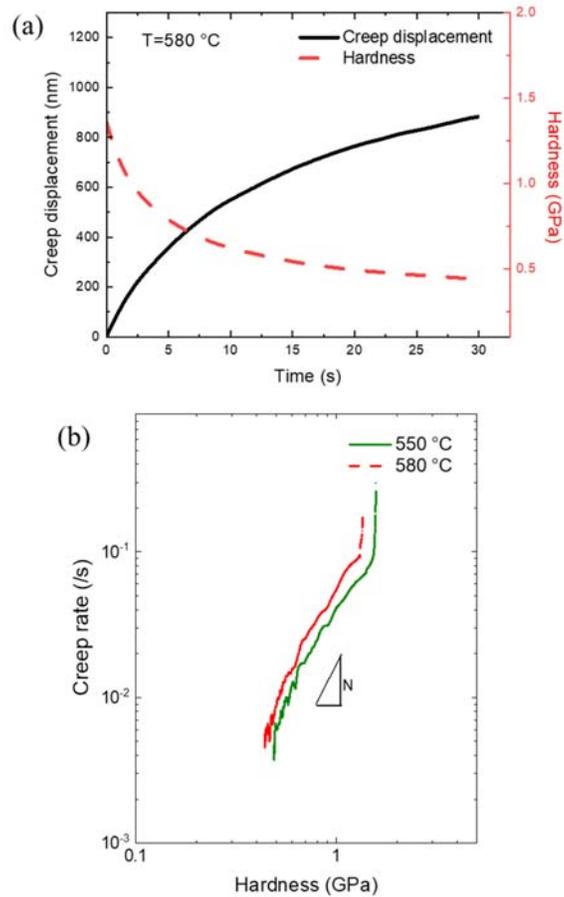

**Fig. S5:** (a) Representative evolution of time-dependent displacement and simultaneous reduction of contact stress or hardness and (b) the change in indentation strain rate or creep rate with hardness during holding at constant load of 9 mN for temperatures above the BDTT. The indentation strain rate was computed as the ratio of indentation velocity, $dh/dt$, to the indentation depth, $h$ [6]. After an initial rapid decrease, the strain rate varies almost linearly with hardness or equivalently the contact stress. The indentation stress exponent, $N$, is calculated from the slope of the linear portion of the curve [6–8].